\documentclass[12pt]{article}

\begin{document}

\begin{center} {\bf Electrical and thermal transport properties in high T$_c$
superconductors : effects of a magnetic field} \vskip 1.0cm

{\large \bf M. Ausloos \footnote{e-mail: ausloos@gw.unipc.ulg.ac.be}} \\ {\it
SUPRAS}, B5 Sart Tilman University Campus, \\ B-4000 Li\`ege, Belgium \vskip
0.6cm \end{center}

{\noindent \bf Abstract} \vskip 1cm

Experimental studies of the electric and heat currents in the normal,
superconducting and mixed states of high T$_c$ superconductors (HTcS) lead to
characterization, complementary to data obtained from equilibrium 
property based
techniques. A magnetic field superimposed on the superconducting 
sample generates
{\it magneto-transport phenomena}, from which an excess electrical resistivity,
an excess thermoelectric power, the Hall or the Nernst effect. Different
behavioral effects allow one to distinguish various dissipation 
mechanisms, like
quasi particle scattering, vortex motion dissipation and superconductivity
fluctuations, in particular when the Corbino geometry is used. Moreover bulk
measurements of the thermal conductivity and the electrothermal 
conductivity in a
magnetic field give us sure indications of the order parameter symmetry. The
location of the mixed state phase transition lines in the technological phase
diagram of HTcS are briefly pointed out through precise measurements performed
over broad temperature and magnetic field  ranges. The results are mainly
reviewed with the aim of defining further investigation lines.

\vskip 1cm keywords: magneto-transport phenomena, mixed state, high critical
temperature  superconductors, electrical resistivity, thermoelectric power,
Nernst effect, Corbino geometry, thermal conductivity, electrothermal
conductivity, order parameter symmetry, phase transition lines

\section{Introduction}

The location of the various mixed state phases in the so-called technological
phase diagram of high T$_c$ superconductors (HTcS)  as well as the behavior of
vortices are of major interest for inventing low temperature devices. Precise
measurements (Sect. II) performed over broad temperature and magnetic field
ranges can be analysed in order to unveil the vortex lattice phases, scattering
processes and phase transition lines. The utility of a Corbino geometry is
recalled in Sect. III.

Experimental studies of the electric and heat currents in the normal,
superconducting and mixed states of HTcS have been made in $SUPRAS$ 
[1] order to
lead to fundamental characterizing parameters. Those properties are 
complementary
to the data obtained from equilibrium property based techniques. In fact, the
magnetic field imposed on the superconducting sample with an 
orientation parallel
to the electric or temperature gradient generates {\it magneto-transport
phenomena},  from which one can deduce an excess electrical resistivity, or an
excess thermoelectric power or an excess thermal conductivity (Sect. IV).
Moreover if the magnetic field has an orientation perpendicular to the electric
or temperature gradient (Sect. V), it generates the Hall and the Nernst effect
respectively. Different features in these effects allow one to distinguish
various dissipation mechanisms, like quasi particle scattering, vortex motion
dissipation and superconductivity fluctuations, phase transition lines in the
mixed state, and the order parameter symmetry as discussed in 
published papers by
the $SUPRAS$ group [1].

\section{Experimental techniques and data}

In the $SUPRAS$ Measurements and Instrumentation Electronics 
Laboratory ($MIEL$)
home made devices and experimental set-ups  are developed in order to get fine
data. In particular we have shown that we can simultaneously measure the
electrical resitivity together with the thermoelectric power [2,3], the
thermoelectric power together with the thermal conductivity [4,5] and the
thermoelectric power, the thermal conductivity and the thermal diffusivity
together in absence or presence of magnetic fields [6]. Confidence in the
validity of the data recalled here below have been taken due to such 
set-ups and
various tests like in [7].

\section{Corbino geometry}

A method, using a Corbino disk sample geometry, is described in refs. 
[8] and has
been applied to study the resistive tails of sintered $YBCO$. When 
the transport
current passes radially from the rim of the disk sample to its center, the two
component potential drop signal $W$ is detected below T$_c$. It is 
assumed to be
due to (i) quasiparticles $W_q$ and (ii) vortex-core-motion $W_{\phi}$  related
contributions. When the contact pairs for $W$ are placed radially, $W_q(r)$ and
$W_{\phi}(r)$ are found to follow markedly distinctive functional dependences,
providing a unique possibility to deconvolute the relative strengths of both
contributions. The results obtained suggest that the mixed state dissipation of
high-T$_c$ superconductors is strongly influenced by the quasiparticle
excitations, - a point of view not often shared or misrepresented. .

\section{Longitudinal effects}

\subsection{Temperature averaging}

Among our on-going investigations about the mixed state we have been 
intrigued by
the various lines used to dissect the ($B,T$) plane. We have shown that a
distribution of vortex lattice melting temperatures is rather usual 
in fact, and
should smear quite a bit the theoretical features.  In the mixed state part of
the technological phase diagram of $Bi$-2212 tapes [9] we have been able to fit
$R(T)$ data over a wide range of temperature and fields. A related 
procedure has
been followed for discussing the vortex lattice melting and viscosity in
$Y_{0.6}Dy_{0.4}Ba_2Cu_3O_{7-x}$ superconductor studied by electrical 
resistivity
measurements [10]. It has been pointed out that the vortex lattice melting
transition should not be confused with the irreversibility line of magnetic
studies nor with the percolation temperature line for electrical resistivity.

\subsection{Activation energy}

An unusual stretched exponential $B$ $exp^{(-B^{1/4})}$ behaviour for 
the excess
thermal resistivity in the temperature range near T$_c/2$ was obtained by
Bougrine et al. [11]; see also ref. [12]. The former authors have derived the
origin of such a theoretical law from a model including bound and free vortices
on intragranular and intergranular defects. The exponent (1/4) was 
shown to be a
particular value specific to the ($B,T$) regime. In fact 
characteristic  lengths
like the mean free path, the penetration depth and the characteristic 
defect size
control the value of the exponent in the streched exponential. In the original
paper, the theory was intended to follow the first line of approach 
proposed for
modificatins of the superconducitng state by the normal state below 
T$_c$, i.e. a
modification of the phonon mean free path. However the paper can be 
(better) read
again considering that the modification is due to the behaviour of electron, or
the quasi-particle, mean free path variation instead.

\subsection{Field dependence of excess quantities}

We have also examined whether some information on the order parameter symmetry
could be obtained from excess quantities, by removing the temperature 
dependence.
It is known that an $s$-wave gap parameter leads to exponential dependence of
properties, but $d$-waves imply power laws. This is known for the temperature
depedence. We have searched for the same effect due to the magnetic field
influence, i.e. mainly in the mixed state, plotting the excess quantity

\begin{equation} {\Delta Q(T,B)} =  Q(T,B) - Q(T,0). \end{equation}

where $Q$ is the electrical resitivity $\rho$ or the Seebeck coeficient $S$ for
example. The subsequent analysis makes uses of the temperature integral of $\
Delta Q(T,B)$, called the {\it area} $A_Q$, between $T_1$ and $T_2$, 
where $T_1$
is taken much below the field dependent percolation temperature and $T_2$ is
taken much above the superconductivity onset temperature. This method has the
advantage of somewhat eliminating extrinsic contributions. Next it also reduces
the necessity of precisely measured data and high calibration. Finally, it
eliminates the temperature and emphasizes the field dependence. We 
have observed
a linear law, i.e. $log (A_Q)$ = $a_Q ~log ~(B) ~+~ b_Q$, and extracted $a_Q$.
The $a_Q$  values can be related to the field dependence appearing in 
the density
of states and the quasi-particle scattering relaxation time 
dependence appearing
in the kinetic definition of transport properties. They depend on whether the
order parameter has a $d$- or $s$-wave symmetry.  A summary of 
exponents is given
in ref. [13]. They markedly indicate simple fractional values related 
to the fact
that the symmetry of the order parmaeter is likely to be of $d$-type.

\subsection{Electrothermal conductivity}

We have looked at the electrothermal conductivity, obtained from the 
ratio of the
thermopower to the electrical resistivity $\rho$ [14,15]. The electrothermal
conductivity is a very convenient property, because it is, in principle,
independent of the vortex viscosity and has nothing to do with vortex 
motion. It
only measures the dissipation of normal quasiparticles, through some $P_{pq}$
term, both inside and outside the vortex core. The electrothermal conductivity
can be measured in the set-up [2] in which we measure simultaneously $\rho$ and
$S$ from 20 K to room temperature. It was of interest to observe the effect in
the vicinity of the critical temperature and in the mixed state.

Working out the $s$-wave and $d$-wave theory for $P_{pq}$, we have used the
resulting theoretical formulae for fitting the experimental data. The 
parameters
coming out from an $s$-wave picture are noticably wrong in the order of
magnitude.  A fit using the $d$-wave theoretical formulae lead to 
very nice fits
and very reasonable values of the parameters. I stress that this is to my
knowledge the only data showing the type of order parameter symmetry 
at this time
in the vicinity of the critical temperature in HTcS.

\subsection{Magnetothermal conductivity}

The influence of a magnetic field on the thermal conductivity is similar in
various high-T$_c$ cuprates: the magneto-thermal conductivity $\kappa(B)$ is
observed to decrease as the magnetic induction B is increased and this relative
decrease is less pronounced when the temperature is raised.  The first
phenomenological model of the magneto-thermal conductivity of high-T$_c$
superconductors assumed that phonons were moving as Bloch waves in a periodic
vortex lattice potential, thus supposing that the vortex lattice is quite
regular. It has been noticed that the influence of a magnetic field on the
transport properties in a field are quite similar in various 
high-T$_c$ cuprates
and the temperature dependence could be well described by rather considering an
electronic origin of the peak observed below $T_c$ [17,18] together 
with assuming
a $d_{x^2-y^2}$-wave gap parameter and Van Hove singularities in the electronic
spectrum [19], - see previous subsections.

We have compared the field dependence of the electronic contribution $\kappa_e$
to the thermal conductivity of a $d$-wave superconductor both for $s$-wave or
$d$-wave order parameter symmetry cases [20-22]. We have 
undoubtlessly shown that
experimental results on various high-T$_c$ materials can be well 
described by the
$d$-wave model though only with a small difference between the deduced physical
parameters from the $s$-wave case in absence of field. However, the theoretical
field dependence of $\kappa_e$ at very low temperature has been shown by Houssa
et al. to be incompatible with our calculations on the magneto-thermal
conductivity of an $s$-wave superconductor. The data on $YBa_ 
2Cu_3O_{7-\delta}$
and $Bi_2Sr_2CaCu_2O_8$ is much better reproduced with a $d$-wave gap parameter
[21,22].

 From several publications discussing data on $Bi$-, $Hg$-, and $RE$- based
superconductors  ($RE$ = rare earth) taken for thermomagnetic effects, it has
been shown that interesting parameter values can be obtained on 
vortex and quasi
particle scatterings in the mixed state of such superconducting ceramics
materials [22].

\section{Transverse effects}

Pekala et al., in several publications [23], have found that the 
Nernst voltage
is an interesting property to be further studied in order to probe  the ($B,T$)
phase diagram and observe the position and temperature dependence of phase
transition field lines (PTFL), like the upper critical field B$_{c2}$ 
line", the
"melting (m) line",  the "irreversibility (i) line", the "glass (g) transition
line", the "electrical resistivity percolation (p) line" and several 
"structural
PTFL" separating regions in which the vortices form a triangular (t) or square
(s) lattice. Below the B$_{c2}$ PTFL, in the so-called liquid phase, 
it has been
recently predicted that a PTFL exists between the critical fluctuation (f)
(Ginzburg-Levanyuk) PTFL and the m-line. The excess thermoelectric power and
Nernst effect can serve to obtain characteristic PTFL in the mixed state by
looking at the singularities, i.e. break in slopes, in quantities 
like $\rho(T)$,
$S(T)$, or $N(T)$ at fixed $B$ values. The number of PTFL is exactly that
predicted from previous reviews [24] plus one from a new theoretical prediction
in ref. [25] in the liquid phase itself. The case of $Bi$-based HTcS 
ceramics has
been used for illustration in the oral presentation and will be discussed
elsewhere in detail [26].

\section{Conclusions}

Part of the discussion in these reports has been centered on our 
$SUPRAS$ way to
get some information on the order parameter symmetry from transport properties
through the analysis of the temperature (integrated or not) excess 
quantity field
dependence in the mixed state.

I stress that bulk measurements of the thermal conductivity and the
electrothermal conductivity in a magnetic field give as sure indications of the
order parameter symmetry as more microscopic techniques, both at low 
temperature
and ){\it even in the vicinity of the critical temperature}. The results can be
reviewed mainly from our work on $Hg$-, $Bi$- and $RE$- based superconductors
with the aim of defining further investigation lines. However much other groups
have contributed to such investigations. We quote them in our publications, but
they are omitted here for lack of space.

These findings result from $SUPRAS$ [1] group role and objectives, set-up as
early as 1988. Nowadays the theoretical and experimental amount of 
data allows us
to pursue more technical problems.

\vskip 0.6cm {\bf Acknowledgments}

\vskip 0.6cm

I greatullly thank R. Nicolsky for inviting me to $ICMC'2000$ in Rio 
de Janeiro ,
and G. Fraga, J. Schaff and P. Pureur for making my stay so easy in 
Porto Alegre,
and Brasil, before and during the meeting. The reported work would 
not have been
possible without the intense activity of the leading members of the $SUPRAS$
group, be they rather permanent or long term visitors, whom I thank, i.e. H.
Bougrine, P. Clippe, R. Cloots, S. Dorbolo, K. Durczewski, M. Houssa, J. Mucha,
S. K. Patapis, M. Pekala, A. Pekalski, A. Rulmont, S. Sergeenkov, Ph.
Vanderbemden, H.W. Vanderschueren and N. Vandewalle (in alphabetical order).

\newpage
[1] see  http://www.supras.phys.ulg.ac.be/

\vskip 0.2cm [2] Ch. Laurent, Ph. D. thesis, 1988, Univ. Li\`ege, 
unpublished; A.
Vanderschueren, H. Bougrine, H.W. Vanderschueren and M. Ausloos, 
Proc. IMEKO TC-4
(CTU Prague-FEE, 1995) p. 362.

\vskip 0.2cm [3] S.K. Patapis, M. Ausloos and Ch. Laurent, in {\it Dynamics of
Magnetic Fluctuations in High Temperature Superconductors}, Ed. by G. 
Reiter, P.
Horsch and G.C. Psaltakis (Plenum, New York, 1991) p. 207; Ch. Laurent, M.
Ausloos et S.K. Patapis, Rev. Phys. Appl. 24 (1989) 501.

\vskip 0.2cm [4] H. Bougrine, Ph. D. thesis, 1994, Univ. Li\`ege, 
unpublished; H.
Bougrine, H.W. Vanderschueren, A. Vanderschueren, and M. Ausloos, Proc. IMEKO
TC-4 (CTU Prague-FEE, 1995) p. 367.

\vskip 0.2cm [5] H. Bougrine and M. Ausloos,  Rev. Sci. Instrum. 66 (1995) 199.

\vskip 0.2cm [6] S. Dorbolo, Ph.D. thesis, 2000, Univ. Li\`ege,  unpublished.

\vskip 0.2cm [7] H. Bougrine and M. Pekala, Supercond. Sci. Technol. 10 (1997)
621.

\vskip 0.2cm [8] V. V. Gridin, S. Sergeenkov and M. Ausloos, Solid 
State Commun.
98 (1996) 623;  S. A. Sergeenkov, V. V. Gridin,  and M. Ausloos,  Z. Phys. B-
Condensed Matter 101 (1996) 565.

\vskip 0.2cm [9]  M. Pekala, H. Bougrine, W. Gadomski, C.G. Morgan, C.R.M.
Grovenor, R. Cloots, and M. Ausloos, Physica C 303 (1998) 169.

\vskip 0.2cm [10] M. Ausloos, in Proc. {\it 11th Seminar of Phase 
Transitions and
Critical Phenomena}, Phys. Chem. Solids Series, M. Kazimierski, Ed. (ILTSR-PAN,
Wroclaw, 1998) p. 44.

\vskip 0.2cm [11] H. Bougrine, S. Sergeenkov, M. Ausloos, and M. Mehbod, Solid
State Commun. 86 (1993) 513.

\vskip 0.2cm [12] S. Sergeenkov and M. Ausloos, Phys. Rev. B 52 (1995) 3614.

\vskip 0.2cm [13]  M. Ausloos, R. Cloots and M. Pekala, J. Supercond. 11 (1998)
515.

\vskip 0.2cm [14] M. Houssa, M. Ausloos, and M. Pekala, Phys. Rev. B 
54 (1996) R
12 713.

\vskip 0.2cm [15] M. Houssa, R. Cloots, S. Stassen, M. Pekala, and M. Ausloos
Czech. J. Phys. 46, S2 (1996) 1003.

\vskip 0.2cm [16] M. Houssa, Ph. D. thesis, Univ. Li\`ege, 1996, unpublished

\vskip 0.2cm [17] M. Houssa, H. Bougrine, M. Ausloos, I. Grandjean, and M.
Mehbod,  J. Phys. Cond. Matter. 6 (1994) 6305

\vskip 0.2cm [18] M. Ausloos and M. Houssa, Physica C 218 (1993) 15; M. Houssa
and M. Ausloos, Phys. Rev. B 51 (1995) 9372.

\vskip 0.2cm [19]  M. Houssa, M. Ausloos, and K. Durczewski, Phys. Rev. B 54
(1996) 6126.

\vskip 0.2cm [20] M. Ausloos and M. Houssa, J. Phys.: Condens. Matter 7 (1995)
L193.

\vskip 0.2cm [21] M. Houssa and M. Ausloos, Z. Phys. B  101 (1996) 
353; M. Houssa
and M. Ausloos, J. Phys. : Condens. Matt. 9, (1997) 201.

\vskip 0.2cm [22] M. Ausloos and M. Houssa, Supercond. Sci. Technol.  12 (1999)
R103.

\vskip 0.2cm [23]  M. Pekala, H. Bougrine,  and M. Ausloos,  J. Phys.: Condens.
Matter. 7 (1995) 5607  (1995); M. Pekala, H. Bougrine, T. Lada, A. 
Morawski  and
M. Ausloos, Supercond. Sci. Technol. 8 (1995) 726; M. Pekala, E. 
Maka, D. Hu, V.
Brabers, and M. Ausloos, Phys. Rev. B 52 (1995) 7647; M. Ausloos, M. Pekala, H.
Bougrine, T. Lada and A. Morawski,  Physica C 252 (1995) 1; M. Pekala and M.
Ausloos, in Proc. NATO ASI on {\it High Temperature Superconductors, 
Physics and
Materials Sciences of Vortex  States, Flux Pinning \& Dynamics}, R. 
Kossowsky, S.
Bose, V. Pan, Z. Durusoy, Eds. (Kluwer, Dordrecht, 1999) p. 559; S. 
Lambrecht and
M.Ausloos,  Phys. Rev. B 53 (1996) 14047.

\vskip 0.2cm

[24] G. Blatter, M.V. Feigel'man, V.B. Geshkenbein, A.I. Larkin and 
V.M. Vinokur,
Rev. Mod. Phys. 66 (1994) 1125; M. Ausloos, Molec. Phys. Rep. 24 (1999) 158.

\vskip 0.2cm [25]  S.K. Chin, K. Nguyen and A. Sudbo, Phys. Rev. B 59 (1999)
14017.

\vskip 0.2cm

[26] M. Ausloos and M. Pekala, in preparation

\end{document}